\documentclass[apj,numberedappendix]{emulateapj}
\usepackage{apjfonts}

\shorttitle{STABLE BURNING ON ACCRETING WHITE DWARFS}
\shortauthors{SHEN \& BILDSTEN}
\slugcomment{Accepted for publication in The Astrophysical Journal}

\newcommand{\be}{\begin{eqnarray}}
\newcommand{\ee}{\end{eqnarray}}
\newcommand{\lp}{\left(}
\newcommand{\rp}{\right)}
\newcommand{\inv}{^{-1}}
\newcommand{\E}[1]{\times10^{#1}}

\begin{document}


\title{Thermally Stable Nuclear Burning on Accreting White Dwarfs}

\author{Ken J. Shen}
\affil{Department of Physics, Broida Hall,\\University of California,
Santa Barbara, CA 93106;\\kenshen@physics.ucsb.edu}

\author{Lars Bildsten}
\affil{Kavli Institute for Theoretical Physics and Department of Physics, Kohn Hall,\\University of California, Santa Barbara, CA 93106;\\bildsten@kitp.ucsb.edu}


\begin{abstract}

One of the challenges to increasing the mass of a white dwarf through accretion is the tendency for the accumulating hydrogen to ignite unstably and potentially trigger mass loss.  It has been known for many years that there is a narrow range of accretion rates for which the hydrogen can burn stably, allowing for the white dwarf mass to increase as a pure helium layer accumulates.  We first review the physics of stable burning, providing a clear explanation for why radiation pressure stabilization leads to a narrow range of accretion rates for stable burning near the Eddington limit, confirming the recent work of Nomoto and collaborators.  We also explore the possibility of stabilization due to a high luminosity from beneath the burning layer.  We then examine the impact of the $\beta$-decay-limited ``hot'' CNO cycle on the stability of burning. Though this plays a significant role for accreting neutron stars, we find that for accreting white dwarfs, it can only increase the range of stably-burning accretion rates for metallicities $<0.01Z_\odot$.

\end{abstract}



\section{Introduction}

The commonly accepted explanation for Type Ia supernovae is the thermonuclear ignition of carbon in the core of a massive white dwarf \citep[see review of][]{hil00}.  The favored channel for mass growth of the white dwarf (WD) is accretion from a companion star, a topic that has been well-studied \citep[e.g.,][]{sien75,pz78,sion79,nom82}.  However, thermal instabilities in the accreting hydrogen/helium layer result in thermonuclear runaways that may inhibit mass growth for accretion rates, $\dot M$, below a critical value of $\sim 10^{-7} \ M_\odot $ yr$\inv$ \citep{pz78,sion79,sien80,fuji82b,pac83,yoon04}.  It is clear that for much lower rates  $<10^{-9} \ M_\odot $ yr$ \inv$, the accumulation of the hydrogen layer always leads to mass ejection during a classical nova event \citep{fuji82a,fuji82b,mac83,gehrz98,tb04}.

Analytic studies of steadily burning and accreting atmospheres on WDs have neglected the $\beta $-limited CNO cycle, whereas studies of neutron stars find that the CNO cycle is strongly limited by $\beta$-decays \citep{fhm81,bild98}. When the rate-limiting step is a $\beta$-decay, the nuclear energy generation rate is independent of temperature, and thus the burning is stable to thermal perturbations. What is not known is the metallicity of the material accreted on the WD needed for the $\beta$-limited CNO cycle to become important during stable burning.  If the inclusion of the full CNO cycle can meaningfully increase the range of accretion rates that lead to thermally stable burning layers on accreting WDs, it would substantially impact our understanding of Type Ia progenitor scenarios.  

We begin in \S\ref{sec:steady} and \S\ref{sec:stab} by rederiving the thermonuclear stability criterion for hydrogen accretion onto a WD and providing an explanation as to why the calculated range of $\dot M$'s for stable burning is so narrow. In \S\ref{sec:fullCNO} we incorporate the full CNO cycle, including the temperature-insensitive $\beta$-decays, and show that this can modify the boundary for stable burning only at low metallicities $Z<0.01Z_\odot$. These metallicities are so low that it is unlikely they are relevant to most progenitor scenarios. We close in \S\ref{sec:conc} by summarizing our work.


\section{Steady-State Hydrogen-Burning Solutions}
\label{sec:steady}

We follow the one-zone formalism of \cite{pac83}, in which quantities such as temperature and pressure have a single value and are defined at the bottom of the hydrogen-burning layer.  We neglect the effect of the underlying helium layer except as a source of luminosity, but see \cite{jose93} for a numerical two-zone study of unstable hydrogen and helium shell flashes on WDs, and \cite{cn06} for a two-zone model of X-ray bursts on neutron stars.  We assume a non-degenerate equation of state, and we include radiation pressure.  We also assume radiative energy transport via Thomson scattering.  These assumptions are justified in Figure \ref{fig:steady}.

By energy conservation in steady-state, the exiting radiative luminosity, $L_0$ (the 0-subscript denotes the steady-state value), must be equal to the sum of the luminosity emitted from the core, $L_{\rm c}$, plus the luminosity from nuclear burning.  It is safe to ignore the luminosity from compression of the accreting material as it is orders of magnitude lower than the luminosity from burning (see \cite{tb04} for an analytic approximation to the compressional luminosity).  In steady-state, material is burned as fast as it is accreted, so
\be
	L_0 = \dot{M}XE + L_{\rm c} ,
	\label{eq:lum}
\ee
where the hydrogen mass fraction is taken to be $X=0.7$ unless otherwise specified, and $ E = 6.4 \E{18} $ erg g$\inv$ is the energy released when a unit mass of hydrogen is converted to helium.  The radiative-zero solution gives the luminosity as
\be
	L = \frac{ 4\pi }{ 3 } \frac{ GMc }{ \kappa_{\rm T} } \frac{ aT^4 }{ P } ,
	\label{eq:radzero}
\ee
where the Thomson opacity is $ \kappa_{\rm T} = \lp 1 + X \rp \sigma_{\rm T} / 2 m_{\rm p} $ and $M$ is the mass of the WD.  The envelope mass, $M_{\rm env}$, is very small and is neglected in comparison with $M$.

The natural unit for the luminosity is the Eddington luminosity,
\be
	L_{\rm Edd} = \frac{ 4\pi GMc }{ \kappa_{\rm T} } .
\ee
Since stable nuclear burning on a WD releases much more energy than gravitational infall, the accretion rate is given in units of the nuclear Eddington accretion rate assuming complete hydrogen depletion,
\be
	\dot{ M }_{\rm Edd} = \frac{ 4\pi GMc }{ \kappa_{\rm T} XE } .
\ee

Equations (\ref{eq:lum}) and (\ref{eq:radzero}) are rewritten in terms of $\beta$, the ratio of gas pressure to total pressure, as
\be
	1-\beta_0 = \dot{a} + f ,
	\label{eq:eddstdmdl}
\ee
where the dimensionless accretion rate and core luminosity are $\dot{a} = \dot{M} / \dot{M}_{\rm Edd}$ and $f=L_{\rm c}/L_{\rm Edd}$, respectively.  Equation (\ref{eq:eddstdmdl}) is an extension of the Eddington standard model \citep[e.g.,][]{hkt04}.  This relation is useful when we discuss the effect of radiation pressure on stability.  Since $ 0 < \beta < 1 $, a necessary condition for steady-state burning is $ \dot{ a } + f < 1 $; that is, the sum of the contributions to the luminosity from accretion and the core cannot exceed the Eddington limit.

The specific entropy, $s$, is governed by
\be
	T \frac{ ds }{ dt } &=& \epsilon - \frac{ \partial L }{ \partial M } \nonumber \\
	&=& \epsilon - \frac{ L - L_{\rm c} }{4\pi R^2} \frac{g}{P} ,
	\label{eq:entr}
\ee
where $\epsilon$ is the energy generation rate, and the burning layer mass has been approximated as the mass contained in a scale height, $h = P / \rho g$.  The gravitational acceleration at the base of the envelope is $ g=GM/R^2$, with $R$ the radius of the WD core.  Using a standard thermodynamic relation, equation (\ref{eq:entr}) is rewritten in the form
\be
	\frac{ dT }{ dt } = \frac{ 1 }{ c_P } \left[ \epsilon - \frac{ L - L_{\rm c} }{ 4\pi R^2} \frac{ g }{ P } \right] + \frac{ T }{ P } \nabla_{\rm ad} \frac{ dP }{ dt } .
	\label{eq:temp}
\ee
The specific heat at constant pressure when non-degenerate is
\be
	c_P = \frac{ 32 - 24\beta - 3\beta^2 }{ 2 \beta^2} \frac{ k_{\rm B} }{ \mu m_{\rm p} } ,
\ee
where $ \mu $ is the mean molecular weight, and
\be
	\nabla_{\rm ad} = \left. \frac{ \partial \ln T }{ \partial \ln P } \right|_s = \frac{ 2 \lp 4 - 3\beta \rp }{ 32 -24\beta -3\beta^2 }
	\label{eq:delad}
\ee
is the adiabatic temperature gradient.  The first term on the right side of equation (\ref{eq:temp}) contains contributions from heating by nuclear burning and core luminosity into the zone, and cooling by radiative luminosity out of the zone.  The second term represents a temperature change due to adiabatic expansion or compression of the accreted matter.

In steady-state, the time-derivatives in equation (\ref{eq:temp}) are equal to zero, so
\be
	\epsilon_0 = \frac{L_0-L_{\rm c}}{4\pi R^2}\frac{g}{P_0} .
\ee
Thus, given $L_{\rm c}$ and $\dot{M}$, we can find steady-state (but not necessarily stable) solutions for $ T_0 $ and $ \rho_0 $, as shown in Figure \ref{fig:steady} for the case of hydrogen-burning in the ``cold'' CNO limit.


\subsection{The Cold CNO Limit}
\label{sec:coldCNO}

Hydrogen burning proceeds via the CNO cycle at the temperatures ($ > 10^7 $ K) relevant to accretion rates of $\dot M>10^{-10} \ M_\odot $ yr$\inv$ \citep{fuji82b,tb04}. The slowest step in the CNO cycle for $ T < 10^8 $ K and $ \rho < 10^4 $ g cm$^{-3}$ is $ ^{ 14 } $N(p,$ \gamma $)$ ^{ 15 } $O, and so nearly all catalytic nuclei are in the form of $ ^{ 14 } $N.  In this ``cold'' CNO limit, the energy generation rate is
\be
	\epsilon_{\rm cold}
		& = & \frac{ Q_{\rm CNO} }{ \rho } \frac{ dn_{14\rm{N}} }{ dt } \nonumber \\
		& = & 4.4 \E{25} \rho X Z_{ \rm CNO} \times \lp \frac{ \exp{ \lp -15.231 / T_9^{ 1/3 } \rp } }{ T_9^{ 2/3 } } \right. \nonumber \\
		&& \left. + 8.3 \E{-5} \frac{ \exp{ \lp - \frac{3.0057}{T_9} \rp } }{ T_9^{ 3/2 } } \rp \textrm{ erg g$\inv$ s$\inv$,}
\ee
where $ Q_{\rm CNO} = 4.0 \E{-5} $ erg is the energy released for one loop of the CNO cycle, the reaction rate for $ ^{ 14 } $N is obtained from \cite{lem06}, and $ T_9 = T/10^9 $ K.  The mass fraction of CNO isotopes is $ Z_{\rm CNO} $, and we assume the total metallicity to be $ Z=3Z_{\rm CNO} $.\footnote{Though there is evidence for enhanced CNO abundances in nova ejecta \citep{gehrz98}, we will presume that the metallicity in the accumulating layer matches that of the accreted material.}  The first term in the parentheses is the non-resonant reaction with an S-factor of 1.7 keV-barn.  The second represents the 259 keV resonant reaction, which dominates for $T \gtrsim 1.5 \E{8} $ K.

\begin{figure}
	\plotone{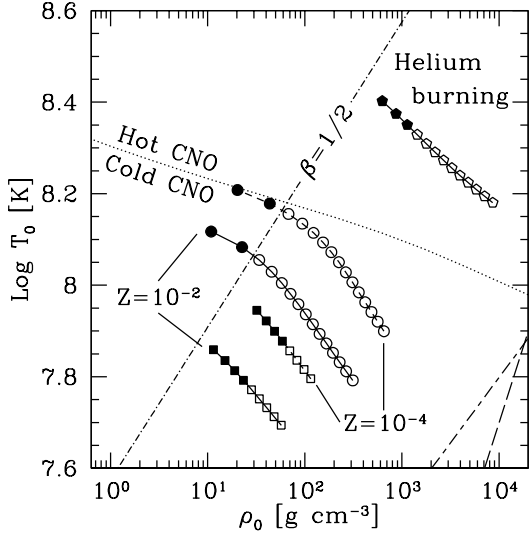}
	\caption{Temperature and density at the burning depth for varying WD mass and accretion rate assuming steady burning.  Shown are WDs burning hydrogen in the cold CNO limit with $M = 0.5 \ M_\odot$ (\emph{squares}) and $ 1.35 \ M_\odot$ (\emph{circles}), with metallicities $Z=10^{-2}$ (\emph{solid lines}) and $10^{-4}$ (\emph{dashed lines}); $Z$ is not given in solar units.  Also shown are solutions for a $0.9 \ M_\odot$ WD accreting pure helium (\emph{pentagons}).  The lower-right-most point of the hydrogen-burning solutions has $\dot{M}=10^{-8}M_{\odot} $ yr$\inv$, while the lower-right-most helium-burning solution has $\dot{M}=10^{-7}M_{\odot} $ yr$\inv$.  Connected points moving towards the upper-left increase logarithmically with 8 points per decade of $\dot{M}$.  Filled points are thermally stable.  Also shown are lines to the left of which the layer is non-degenerate (\emph{long-dashed line}), opacity is due to Thomson scattering (\emph{long-dashed-short-dashed line}), and radiation pressure dominates (\emph{dashed-dotted line}), and below which hydrogen burns in the cold CNO limit (\emph{dotted line}).}
	\label{fig:steady}
\end{figure}

Shown in Figure \ref{fig:steady} are the locations of the steadily-burning hydrogen layer for WDs of mass $0.5 \ M_\odot$ (\emph{squares}) and $1.35 \ M_\odot$ (\emph{circles}), accreting matter at $\dot{M} \geq 10^{-8} \ M_\odot$ yr$\inv$ with metallicities $Z=10^{-2}$ (\emph{solid lines}) and $Z=10^{-4}$ (\emph{dashed lines}).  The hydrogen is assumed to burn in the cold CNO limit.  Filled points are thermally stable (see \S\ref{sec:stab}).  To the left of the long-dashed and long-dashed-short-dashed lines, the layer is non-degenerate and opacity is set by Thomson scattering, respectively.  The steady-state solutions are well within this region, so our approximations are valid.  Earlier examples of these solutions can be found in \cite{sien75} and \cite{nom06}.  For completeness, we also show the location where steady-state helium burning (\S\ref{sec:helium}) would occur for a $0.9 \ M_\odot$ WD (\emph{pentagons}).

A few trends are clear.  For a fixed $\dot{M}$, increasing the WD mass, and thus $g$, drives the steady-state temperature and density higher.  The dotted line shows where the $\rm{p}+^{14}$N reaction timescale is equal to the $^{15}$O $\beta$-decay time of 176 s; above this curve the CNO cycle is $\beta$-limited.  At low metallicities of $10^{-2}Z_\odot$ (\emph{dashed lines}), the layer is hot enough that the $\beta$-limited CNO cycle is non-negligible for the highest accretion rates and most massive WDs. Since these are the WDs of direct interest to the Type Ia progenitor problem, we pursue this question further in \S\ref{sec:fullCNO}.  However, we first wish to carry out the analysis needed to determine the thermal stability of the steadily-burning solutions.


\section{Thermal Stability of the Shell Source}
\label{sec:stab}

We begin by perturbing the temperature around its steady-state value via
\be
	T(t) = T_0 + \delta T \, \textrm{e}^{ \, t / \tau } \textrm{,}
\ee
where $ \delta T / T_0 \ll 1 $.  If the perturbation timescale, $ \left| \tau \right| $, is  shorter than the accretion timescale, $t_{\rm acc} = M_{\rm env}/\dot{M}$, then $M_{\rm env}$ can be approximated as constant during the perturbation.  With this assumption, the accompanying pressure perturbation can be related to the density perturbation by the parametrization
\be
	\alpha = \left. \frac{ \partial \ln P }{ \partial \ln \rho } \right|_{M_{\rm env}} ,
\ee
as shown below.  In our final results, the inequality $ \left| \tau \right| \ll t_{\rm acc} $ is indeed satisfied for essentially the entire parameter regime, so this assumption is self-consistent.


\subsection{Relations among Temperature, Pressure, and Density Perturbations for a Constant Mass Envelope}

For a constant luminosity envelope with Thomson opacity, $P(r) \propto T(r)^4$, so $\rho(r) \propto P(r)^{3/4}$, neglecting degeneracy.  Since the envelope mass is much smaller than that of the WD, integration of hydrostatic equilibrium yields the density profile
\be
	\rho(r) = \rho_{\rm b} \left[ 1 - \frac{1}{x} \lp 1 - \frac{R_{\rm b}}{r} \rp \right]^3 \textrm{,}
\ee
where $x=4h/R_{\rm b}$, and the scale height is $h=P_{\rm b}/\rho_{\rm b} g$.  (To avoid confusion,  in this section we explicitly label variables at the bottom of the layer with the subscript ``b''.)  The outer radius of the envelope, where $\rho(R_{\rm o})=0$, is
\be
	R_{\rm o} = R_{\rm b} / (1-x).
\ee
Thus, the scale height is constrained to be $ h < R_{\rm b}/4 $.  The thin solid lines in Figures \ref{fig:stab1} and \ref{fig:stab2} are accretion rates above which this constraint is not met, i.e., where hydrostatic envelope solutions do not exist.  This is a stronger bound than the nuclear Eddington limit (\emph{dotted line in Figure \ref{fig:stab1}}).  This hydrostatic constraint is similar to the ``red-giant luminosity'' \citep{pac70,fuji82b,hknu99} above which the envelope expands to red-giant dimensions and outflow may occur.  The numerical results of \cite{nom06} for this upper bound are shown in Figure \ref{fig:stab1} (\emph{thin dashed line}).  The discrepancy at lower WD masses is likely due in part to the approximation of the burning layer as the mass within a scale height.

The mass of the envelope,
\be
	M_{\rm env} &=& \int_{R_{\rm b}}^{R_o} 4 \pi r^2 \rho(r) dr \nonumber \\
	&=& 4\pi \rho_{\rm b} R_{\rm b}^3 \frac{1}{x^3} \left[ \ln \lp \frac{1}{1-x} \rp -x - \frac{x^2}{2} - \frac{x^3}{3} \right],
\ee
is conserved during the perturbation, so $\delta M_{\rm env} = 0$.  Thus,
\be
	\frac{ \delta \ln \rho_{\rm b} }{ \delta \ln x } &=& \left[ 3 - \lp \frac{x^4}{1-x} \rp \lp \ln  \lp \frac{1}{1-x} \rp - x - \frac{x^2}{2} - \frac{x^3}{3} \rp \inv \right] \nonumber \\
	&\equiv& D.
\ee
From the definition of $x$, we have $\delta \ln x = \delta \ln P_{\rm b} - \delta \ln \rho_{\rm b}$, so $\delta \ln P_{\rm b} = (1 + 1/D) \delta \ln \rho_{\rm b}$, and $\alpha = (1+1/D)$.  Thus, $\alpha$ is completely specified by the steady-state burning conditions.  Note that $x<1$ constrains $\alpha$ to be $<1$ as well.

The equation of state yields the density perturbation in terms of the temperature perturbation:
\be
	\delta \ln \rho_{\rm b} &=& \frac{ \lp \partial \ln \rho / \partial \ln T \rp_P }{ 1 - \alpha \lp \partial \ln \rho / \partial \ln P \rp_T } \delta \ln T_{\rm b} .
\ee
For a non-degenerate gas, $\lp \partial \ln \rho / \partial \ln P \rp_T = 1 / \beta$ and $\lp \partial \ln \rho / \partial \ln T \rp_P = 3 - 4/\beta$, so
\be
	\delta \ln \rho_{\rm b} = - \lp \frac{ 4/\beta-3 }{ 1 - \alpha /\beta } \rp \delta \ln T_{\rm b} .
\ee


\subsection{Thermal Stability Condition}
\label{sec:stabcond}

We apply these perturbations to equation (\ref{eq:temp}).  The steady-state solutions will be stable if $ \mathbb{R}[\tau] < 0 $.  Keeping only first-order terms, we arrive at the condition for stability,
\be
	\frac{A}{B} < 0 ,
	\label{eq:stab}
\ee
where
\be
	A &=& \nu - 4 \lp 1+\frac{f}{\dot{a}} \rp - \frac{4 - 3\beta}{ 1 - \alpha/\beta } \left[ \frac{\lambda}{\beta} + \frac{\alpha}{\beta} \lp 2+\frac{f}{\dot{a}} \rp \right] \nonumber \\
	B &=& 1 + \nabla_{\rm ad} \lp \frac{\alpha}{\beta} \rp \frac{4-3\beta}{1-\alpha/\beta} .
	\label{eq:stabdef}
\ee
The logarithmic derivatives of $\epsilon$ are
\be
	\lambda \equiv \left. \frac{ \partial\ln{\epsilon} }{ \partial\ln{\rho} } \right|_T \qquad
	\nu \equiv \left. \frac{ \partial \ln \epsilon }{ \partial \ln T } \right|_\rho \textrm{.}
\ee

Stability can be achieved via two physically distinct mechanisms, when either $A$ or $B$ is negative.  The case of $A<0$ (the thick solid line in Fig. \ref{fig:stab1} is $A=0$) is the condition where an increase in temperature causes a larger loss of energy due to exiting radiation than gain from energy generation.  This allows for stability when the shell is thin ($h \ll R$).  The second stability mechanism involves the parameter $B$, which is the ratio of gravothermal specific heat, $c_*$, to $c_P$.  Since $c_P>0$, the inequality $B<0$ is equivalent to $c_*<0$.  Physically, this is a situation in which the addition of heat to the shell causes a loss of internal energy due to expansion work \citep{kw90}.  Note that the layer is unstable if both $A$ and $B$ are negative, but, with the assumption that $\lambda$ and $\nu$ are non-negative, this case is ruled out.

Increasing the fraction of radiation pressure (thereby decreasing $\beta$) makes the layer more stable: this is radiation pressure stabilization \citep{sugi78,fuji82b,yoon04}.  In broad terms, \emph{if the plasma is radiation pressure dominated, a small increase in temperature with constant or decreasing pressure will cause the density to decrease dramatically.}  This leads to a large drop in energy generation as well as a large amount of expansion work, both of which quench a thermal instability.  From equation (\ref{eq:eddstdmdl}), decreasing $\beta$ can be achieved by increasing the accretion rate and/or core luminosity.  This helps to explain the requirement of high accretion rates and/or very hot WDs for stability.  


\subsection{The Cold CNO Limit}

\begin{figure}
	\plotone{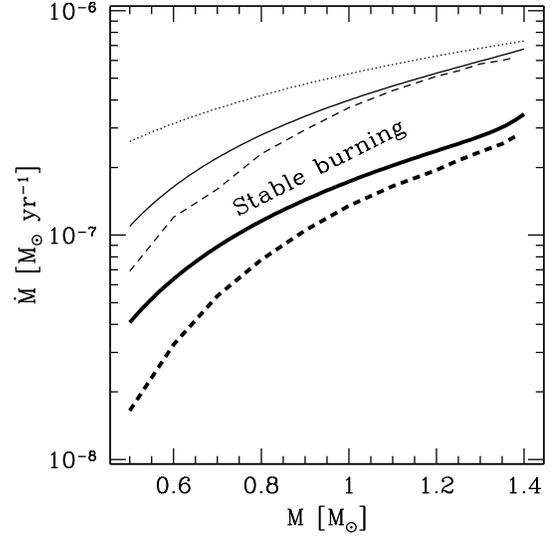}
	\caption{Range of accretion rates for which the hydrogen-burning layer is thermally stable, assuming $Z=10^{-2}$ and no core luminosity.  Above the thin solid line, hydrostatic envelope solutions do not exist.  Below the thick solid line, the atmosphere is thermally unstable.  Stable burning is only possible between the two solid lines.  Also shown are equivalent bounds (\emph{dashed lines}) from a recent numerical simulation \citep{nom06}, as well as the nuclear Eddington limit (\emph{dotted line}).}
	\label{fig:stab1}
\end{figure}

The thick solid line in Figure \ref{fig:stab1} is the minimum accretion rate (i.e., $A=0$ in eqn. (\ref{eq:stab})) for which the hydrogen-burning layer is thermally stable as a function of $M$, assuming $Z=10^{-2}$ and no core luminosity.  The cold CNO limit is not assumed, although it is still applicable.  The slight upturn in the thick solid line for high masses is due to the 259 keV resonance in the $^{14}$N proton capture, which begins to become non-negligible near $T=1.5 \E{8} $ K.  The nuclear Eddington limit is shown for reference (\emph{dotted line}).  Nearly independent of mass, the lowest stable accretion rate is $ \simeq 1/3 $ the limiting accretion rate above which no hydrostatic envelope solutions are found (\emph{thin solid line}).  Also plotted are equivalent bounds (\emph{dashed lines}) from \cite{nom06}.  We now derive the near constancy of this narrow range of thermally stable accretion rates.

The upper bound (\emph{thin solid line}) is where $x=1$, which gives
\be
	\frac{ 4k_{\rm B} T_{\rm max} / \mu m_{\rm p} }{ GM / R} = \beta_{\rm max} ,
	\label{eq:upper}
\ee
where the subscript ``max'' refers to the quantity at the maximum accretion rate.  At the lower bound of stability (\emph{thick solid line}), $h \ll R/4$, and we expand $\alpha$ in powers of $h/R$ to obtain
\be
	\alpha \simeq \frac{16}{5} \frac{h}{R} .
\ee
Expanding equation (\ref{eq:stabdef}) in terms of $h/R$, assuming cold CNO burning and no core luminosity, and only keeping the first order term, yields the condition at the lower bound,
\be
	\nu = 1 + \frac{4}{\beta_{\rm min}} + \frac{16}{5} \frac{ k_{\rm B} T_{\rm min}/\mu m_{\rm p} } { \beta_{\rm min} GM/R} \lp 2+\frac{1}{\beta_{\rm min}} \rp \lp \frac{4}{\beta_{\rm min}}-3 \rp ,
\ee
where the ``min'' subscript refers to the value at the minimum accretion rate for stability.  We have also set $\lambda=1$, which is appropriate for the cold CNO limit.  From Figure \ref{fig:steady}, we see that the temperature does not change significantly over the range of stable accretion, so we approximate $T_{\rm min}=T_{\rm max}$.  Then, equation (\ref{eq:upper}) gives
\be
	\nu = 1 + \frac{4}{\beta_{\rm min}} + \frac{4}{5} \frac{ \beta_{\rm max} } { \beta_{\rm min} } \lp 2+\frac{1}{\beta_{\rm min}} \rp \lp \frac{4}{\beta_{\rm min}}-3 \rp .
\ee
Solving for the ratio of accretion rates yields
\be
	\frac{ \dot{M}_{\rm max} }{ \dot{M}_{\rm min} } = \frac{1}{1-\beta_{\rm min}} \lp 1 - \frac{5\beta_{\rm min}^2}{4} \frac{ \nu - 1 - 4/\beta_{\rm min} }{ \nu + 4 - 6\beta_{\rm min}} \rp .
\ee
Finally, we expand this ratio around $\beta_{\rm min}=3/4$ and, given that the logarithmic temperature derivative of the cold CNO cycle is
\be
	\nu = \frac{ 5.077 }{T_9^{1/3} }  - \frac{2}{3}
\ee
for $T \lesssim 1.5 \E{8} $ K, we assume $\nu=10$ to obtain
\be
	\frac{ \dot{M}_{\rm max} }{ \dot{M}_{\rm min} } = 2.9 \left[ 1 + 2.0\lp \beta_{\rm min} -\frac{3}{4} \rp + O \lp \beta_{\rm min}-\frac{3}{4} \rp^2 \right]
\ee
The coefficient in front varies from $3.1-2.6$ for $\nu=9-12$, and since $\beta_{\rm min}$ never falls below $1/2$, the ratio of accretion rates is tightly constrained to be $\simeq 3$, regardless of mass.  This is the reason for the nearly constant separation between the two solid lines in Figure \ref{fig:stab1}.


\subsection{Importance of Including the Pressure Perturbation}

To simplify the preceding analysis, one might be tempted to assume a constant pressure perturbation ($\alpha=0$) because the layer is ``thin'' -- i.e., $h \ll R$ -- for the minimum thermally stable accretion rate.  However, this will result in significant errors, as we now demonstrate.

Expanding equation (\ref{eq:stabdef}) in terms of $\alpha$ and only keeping the first order term yields the revised stability condition
\be
	\nu - 4 \lp 1+\frac{f}{\dot{a}} \rp - \lp \frac{4}{\beta}-3 \rp \left[ \lambda +\lp 2+\frac{f}{\dot{a}} \rp \alpha \right] < 0,
\ee
or, with the assumption of cold CNO burning and negligible core luminosity,
\be
	\nu < 4+ \lp \frac{4}{\beta}-3 \rp + 2\alpha \lp \frac{4}{\beta}-3 \rp.
\ee
Thus, even if the scale height is only 5\% of the WD radius for $\beta \simeq 1/2$, allowing a pressure perturbation changes the stability condition from $\nu<9$ to $\nu < 10.5$.  Given the strong temperature dependence of the cold CNO cycle, this is a significant change in the minimum thermally stable accretion rate.  For example, a $1.0 \ M_\odot$ WD can accrete stably down to $1.7 \E{-7} \ M_\odot$ yr$\inv$, but if we incorrectly assume a constant pressure perturbation, the minimum accretion rate is $3.0 \E{-7} \ M_\odot$ yr$\inv$, nearly a factor of 2 higher.


\subsection{Pure Helium Accretion}
\label{sec:helium}

In the case of a WD accreting pure helium, the fuel burns via the triple-$\alpha$ reaction, given in \cite{hkt04} as
\be
	\epsilon_{3\alpha} = \frac{5.1 \E{8} \rho^2 Y^3}{ T_9^3} \exp(-4.4027/T_9) \textrm{ erg g$\inv$ s$\inv$,}
\ee
so $ \lambda = 2 $, and
\be
	\nu_{3\alpha} = \frac{4.4027}{T_9} - 3 \textrm{.}
\ee
See Figure \ref{fig:steady} for steady-state conditions of a $0.9 M_{\odot}$ helium-accreting WD (\emph{pentagons}).  For this mass, the range of stable accretion rates is $ 1.7-4.0 \E{-6} \ M_\odot$ yr$\inv$.


\subsection{Flux Stabilization}

\begin{figure}
	\plotone{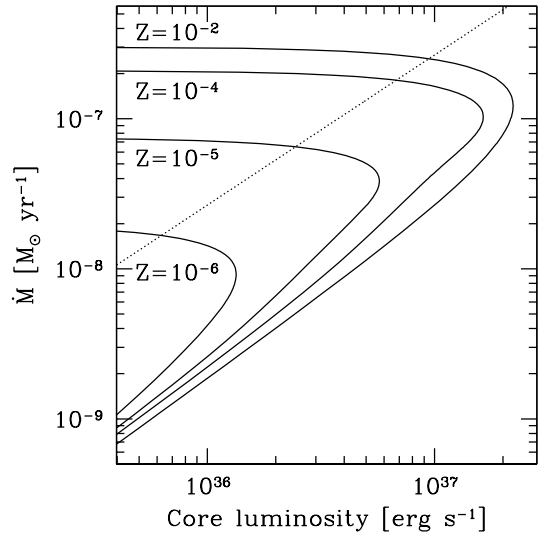}
	\caption{Critical accretion rates vs. core luminosity for a $ 1.35 M_{\odot} $ WD with the given metallicities (\emph{solid lines}).  The atmosphere is thermally stable to the right of these curves.  Also plotted is the luminosity from steady burning of the helium below the hydrogen layer (\emph{dotted line}).  For a given accretion rate, this line represents the maximum possible core luminosity for hydrogen accretion.}
	\label{fig:mdotflux}
\end{figure}

If the WD interior is hot and emits a core luminosity comparable to that from the nuclear energy release of the accreted material, the burning can be stabilized \citep{ft82,pac83}. The regions to the right of the solid curves in Figure \ref{fig:mdotflux} are stabilized by this mechanism. For example, a $1.35 \ M_\odot$ WD with $L_{\rm c}=10^{36}$ erg s$\inv$ will stably burn accreting material at a rate of $10^{-9} \ M_\odot$ yr$\inv$, orders of magnitude lower than the cases presented earlier.  As the core luminosity increases, a larger range of accretion rates are stable.   Our results differ from those of \cite{pac83} due to his assumption of a thin shell ($P=M_{\rm env} g / 4 \pi R^2$) and lack of inclusion of the full CNO cycle, among other reasons.

While a large core luminosity seems an attractive possibility for stabilizing the burning layer, it is difficult to identify an energy source that would provide enough luminosity to affect stability during the accretion of $ \gtrsim 0.1 M_{\odot} $ typically needed to ignite the carbon in the core for a Type Ia supernova.  The most obvious option, the luminosity from the cooling WD itself, is orders of magnitude too small.  Even steady burning of the helium below the hydrogen layer is inadequate to significantly increase the stable accretion rate range.  Complete and stable burning of the helium gives a luminosity, $ L = \dot{M}E_{\rm He} $ (\emph{dotted line in Fig. \ref{fig:mdotflux}}), where $E_{\rm He}=5.8 \E{17}$ erg g$\inv$.  For a $ 1.35 M_{\odot} $ WD, this additional luminosity only lowers the minimum stable accretion rate by $ \simeq 10\% $.

Recent work by \citet{starr04} finds stable hydrogen burning on a hot ($ L \simeq 10^{35}$ erg s$\inv$) accreting WD of $ M = 1.35 \ M_\odot $ for accretion rates down to $ 1.6 \E{-9} \ M_\odot$ yr$\inv$.  Figure \ref{fig:mdotflux} shows that this high luminosity is not responsible for stabilizing the hydrogen burning.  The physics that stabilizes their models at such low accretion rates remains unclear and is at odds with the more recent efforts of \citet{nom06}.


\section{The Full CNO Cycle}
\label{sec:fullCNO}

If the temperature and density at the burning depth are very high, all fusion reactions in the CNO cycle occur more rapidly than the $ \beta $-decays.  The energy generation rate is then determined by the $ \beta $-decays of $^{14}$O and $^{15}$O, which are independent of temperature and density, giving $ \epsilon_{\rm hot} = 5.8 \E{15} Z_{\rm CNO} $ erg g$\inv$ s$\inv$.  (Even for the lowest metallicity cases we consider where $ Z = 10^{-6} $, p-p burning is still negligible for the relevant accretion rates.)  In this $\beta$-limited CNO cycle, hydrogen burning is stable to temperature perturbations.  We now derive the conditions for which the temperature-independent $\beta$-decays of the full CNO cycle can stabilize the burning layer.

The energy generation rate of the full reaction rate network is
\be
	\epsilon = \frac{ \sum_{ ij} Q_{ ij } n_i / t_{ ij } }{ \rho } \textrm{,}
\ee
where $ i $ refers to the isotope, $ j $ differentiates between fusion and $ \beta $-decay reactions for a given isotope, and $ Q_{ij} $ is the energy released in the specified reaction.  In particular, the CNO cycle has a fork at $ ^{ 13 } $N; when cold and/or diffuse, $^{13}$N nuclei preferentially undergo a $\beta$-decay to $^{13}$C, and when hot and/or dense, they capture a proton to become $^{14}$O.  The nuclear reaction timescales, $ t_{ij} $, are defined to satisfy the differential equations
\be
	\frac{ dn_i }{ dt } = \sum_k \frac{ n_k }{ t_k } - \sum_j \frac{ n_i }{ t_{ ij } } \textrm{,}
\ee
where $ k $ refers to the isotopes that can produce isotope $ i $.  With this definition of nuclear reaction timescales, the proton capture timescales are given as
\be
	t_{\rm capture} = \frac{ 1 }{ n_{\rm p} \langle \sigma v \rangle } = \frac{ m_{\rm p} }{ X \rho } \frac{ 1 }{ \langle \sigma v \rangle} \textrm{.}
\ee
The reaction rates, $\langle\sigma{v}\rangle$, are only functions of temperature and are obtained from NACRE \citep{nacre} and LUNA \citep{lem06}.

If the thermal perturbations occur on timescales much longer than reaction timescales, the reaction rate network is always in equilibrium for a given temperature and pressure.  The time for a given nucleus to go once around the CNO cycle is
\be
	t_{\rm cycle} \simeq \frac{Q_{\rm CNO} Z_{\rm CNO}}{14 m_{\rm p} \epsilon} \textrm{,}
\ee
where we have chosen the mean molecular weight of the catalytic nuclei to be $14m_{\rm p}$.  We find that this timescale is always much shorter than the perturbation timescale for all metallicities and thermally stable accretion rates.  Thus, $ dn_i / dt = 0 $.  Then, at a given temperature and pressure, we can solve for all abundances in terms of one of the abundances, such as $ n_{ 12 \rm{ C } } $.  Defining the weighting factor,
\be
	f_{ ij }(T, \rho) \equiv \frac{ n_i / t_{ ij } }{ n_{ 12 \rm{ C } } / t_{ 12 \rm{ C } } } \textrm{,}
\ee
we have
\be
	\epsilon =  \frac{ X n_{12\rm{C}} \langle \sigma v \rangle_{ 12 \rm{ C } } }{ m_{\rm p} } \sum_{ ij } Q_{ij } f_{ ij }(T, \rho) \textrm{.}
\ee
Defining another weighting factor, $ h_i(T,\rho) \equiv n_i / n_{ 12 \rm{ C } } $, yields
\be
	\epsilon
		& = & \frac{ X Z_{\rm CNO} }{ m_{\rm p}^2 } \rho \langle \sigma v \rangle_{ 12 \rm{ C } } \frac{ \sum_{ ij } Q_{ij } f_{ ij }(T,\rho) }{ \sum_i A_i h_i(T,\rho) } \nonumber \\
		& = & X Z_{\rm CNO} G(T,\rho) \textrm{.}
	\label{eq:epfull}
\ee
Here, $ A_i $ is the atomic mass of the isotope, and $ G(T,\rho) $ is a function that depends only on the temperature and density.  This implies that $\nu$ is also solely a function of temperature and density.

With the analytic form of the energy generation rate, finding the steady-state solutions is as before.  Specifying the accretion rate, the core luminosity, the CNO metallicity, and the WD mass fixes the temperature and density at steady-state.  These in turn fix the value of $\nu$.  Thus, given $ L_{\rm c} $, $ Z_{\rm CNO}$, and $ M $, we can analytically find values of $\dot{M}$ for which the atmosphere will satisfy equation (\ref{eq:stab}) and thus be thermally stable.

\begin{figure}
	\plotone{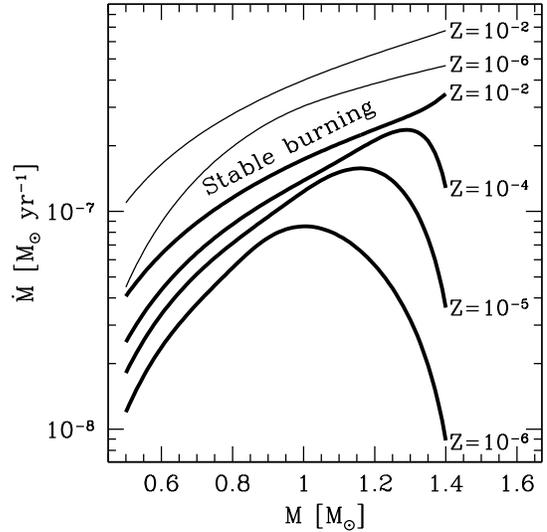}
	\caption{Ranges of thermally stable accretion rates assuming no core luminosity, as in Figure \ref{fig:stab1}, but with the given metallicities.  Burning is via the full CNO cycle.  The upper bound is the rate above which hydrostatic solutions do not exist (\emph{thin solid lines}).  The location of the upper bound has a weak dependence on metallicity; upper bounds for metallicities between solar and $Z=10^{-6}$ lie between the thin solid lines shown.  The lower bound is the transition between thermally stable and unstable burning layers (\emph{thick solid lines}).}
	\label{fig:stab2}
\end{figure}

Metallicity-dependent ranges of accretion rates for which the layer is thermally stable, with burning via the full CNO cycle, are shown in Figure \ref{fig:stab2}.  For accretion rates above the thin solid lines, hydrostatic envelope solutions do not exist, whereas below the thick solid lines, thermal perturbations in the atmosphere grow exponentially.  Lowering the metallicity of the accreting material and increasing the WD mass increase the temperature and density of the burning layer and lead to shorter timescales for proton captures.  This lowers the minimum accretion rate for thermal stability because $ \nu $ decreases as the temperature-independent $\beta$-decays of $^{14}$O and $^{15}$O become more important to energy generation.  This effect is also evident in Figure \ref{fig:mdotflux}.  The limiting case of hot CNO stabilization occurs in the hydrogen-burning shells of neutron stars, where burning proceeds via the $\beta$-limited CNO cycle for solar metallicities at very sub-Eddington accretion rates \citep{bild98} due to the high gravity.


\section{Conclusions}
\label{sec:conc}

We have confirmed that stable burning of hydrogen on an accreting WD requires accretion rates within a narrow range for solar metallicities (see Fig. \ref{fig:stab1}) and shown that this is a robust result set by radiation pressure stabilization.  Increasing the range of thermally stable accretion rates can be temporarily accomplished by having a luminosity exiting the WD core of a magnitude comparable to that released from nuclear burning of hydrogen at that accretion rate (see Fig. \ref{fig:mdotflux}).  However, there is no energy source available to the deep core of the WD that would allow for such stabilization during accretion of the $0.1-0.2 \ M_\odot$ of material needed to reach core ignition for a Type Ia supernovae. Thus, the challenge of a narrow accretion rate regime for stable hydrogen burning of solar metallicity material remains.

In the hopes of alleviating this bottleneck, we considered the stabilizing effect of $\beta$-limited burning of hydrogen via the ``hot'' CNO cycle. This effect reduces the accretion rate needed for stable burning, but requires metallicities of $<10^{-2} \ Z_\odot $ before it opens up much parameter space (see Fig. \ref{fig:stab2}). Such metallicities are relevant to a few globular clusters in our galaxy and others, and for this reason, calculations such as those done by \citet{nom06} should be extended to lower metallicities as they might prove relevant to observations of accreting sources in globular clusters.

However, while very low metallicities can be found in these systems, we know \citep{trem04,lee06} that the metallicity for most of the stellar mass in the nearby universe is larger than $10^{-2} \ Z_\odot$, making it appear unlikely that low-metallicity environments will play a significant role in the overall Type Ia supernova rate.  Full resolution of the impact of the hot CNO cycle awaits the combination of multi-zone calculations with a population synthesis code capable of assessing the contribution of stable burning at lower accretion rates to the Type Ia rate.

We thank the referee for comments and queries that improved the paper.  We also thank Philip Chang, Ken'ichi Nomoto, and Tony Piro for discussions.  This research was supported by the National Science Foundation (NSF) through grants AST 02-05956, PHY 99-07949 and PHY 02-16783 (Joint Institute for Nuclear Astrophysics). 



\end{document}